\documentclass{INTERSPEECH2023}
\usepackage{cite}
\usepackage{amsmath,amssymb,amsfonts}
\usepackage{algorithmic}
\usepackage{graphicx}
\usepackage{textcomp}
\usepackage{xcolor}
\usepackage{booktabs}
\usepackage{hyperref}
\usepackage{makecell}
\usepackage{multirow}
\usepackage[caption=false]{subfig}
\usepackage{float}
\usepackage{hyperref}

\interspeechcameraready

\title{Anomalous Sound Detection Using Self-Attention-Based Frequency Pattern Analysis of Machine Sounds}
\name{Hejing Zhang$^1$, Jian Guan$^{1,*}$\thanks{*Corresponding author}, Qiaoxi Zhu$^2$, Feiyang Xiao$^1$, Youde Liu$^3$}

\address{
$^1$ Group of Intelligent Signal Processing, Harbin Engineering University, China\\
$^2$ Centre for Audio, Acoustics and Vibration, University of Technology Sydney, Australia\\
$^3$ School of Computer Science and Technology, Harbin Institute of Technology, China}
\email{\{zhanghejing, j.guan\}@hrbeu.edu.cn; qiaoxi.zhu@gmail.com; xiaofeiyang128@gmail.com; liuyoudedl@163.com}

\begin{document}

\maketitle
 
\begin{abstract}

Different machines can exhibit diverse frequency patterns in their emitted sound. This feature has been recently explored in anomaly sound detection and reached state-of-the-art performance. However, existing methods rely on the manual or empirical determination of the frequency filter by observing the effective frequency range in the training data, which may be impractical for general application. This paper proposes an anomalous sound detection method using self-attention-based frequency pattern analysis and spectral-temporal information fusion. Our experiments demonstrate that the self-attention module automatically and adaptively analyses the effective frequencies of a machine sound and enhances that information in the spectral feature representation. With spectral-temporal information fusion, the obtained audio feature eventually improves the anomaly detection performance on the DCASE 2020 Challenge Task 2 dataset.

\end{abstract}

\section{Introduction}

Anomalous sound detection (ASD) aims to automatically determine whether the state of a target object is normal or anomalous by analyzing the sound emitted by the object~\cite{Koizumi2020, Kawaguchi2021, Mai2022, suefusa2020anomalous, zeng2022robust, park2018fast, oh2018residual}. ASD is commonly an unsupervised task due to the infrequent and varied occurrence of anomalous machine sounds in real-world scenarios \cite{Koizumi2020, oh2018residual, park2018fast, zeng2022robust, liu2022anomalous, Mai2022}. Therefore, only normal sounds are employed for training to learn the audio feature distribution of normal sounds. The distance between the test sound and the learned normal sound distribution is calculated to detect the anomalous sound having a distance value larger than a threshold \cite{GuanHEU2022, KuroyanagiNU-HDL2022, WeiHEU2022}. 

As an unsupervised task, ASD learns the feature of normal sounds to detect anomalous sounds. If the learnt feature also fits with the anomalous sound, the effectiveness of anomaly detection could be limited. Log-Mel spectrogram has been widely used as the input feature of the machine sound in ASD methods, such as \cite{suefusa2020anomalous, Kapka2020, Giri2020a, dohi2021flow}. However,  in our previous work \cite{liu2022anomalous}, we found that using Log-Mel spectrogram as the audio feature can be ineffective in distinguishing normal and anomalies, as it might filter out high-frequency components of anomaly sound, where distinct features may exist. So spectral-temporal information fusion (STgram) as the audio feature was proposed in \cite{liu2022anomalous}, utilising both the Log-Mel spectrogram and temporal feature extracted from machine sounds. Using STgram, the STgram-MFN method was developed for ASD \cite{liu2022anomalous}, which achieved state-of-the-art performance on the Detection and Classification of Acoustic Scenes and Events (DCASE) Challenge 2020 Task 2 dataset.  

Further investigation indicates that some machine types exhibit prominent characteristics in high frequencies, as evidenced by analyzing the spectrum of machine sounds \cite{zeng2022robust, Mai2022}. Additionally, the models used for anomaly detection in ASD rely heavily on higher frequencies to distinguish between normal and abnormal sounds \cite{Mai2022}. To obtain the audio feature, a high-pass filter is applied before passing it through the Mel filter in \cite{zeng2022robust}, which ranked top 1 in DCASE 2022 Challenge Task 2. The results of experiments demonstrate that this pre-processed feature improved anomaly detection for several machine types, including ToyCar, ToyTrain, Gearbox, and Valve. However, this pre-processing technique \cite{Mai2022, zeng2022robust} relies on the manual or empirical determination of the high-pass filter by observing the effective frequency range in the training data. This approach may be imprecise or time-consuming when implementing ASD in real-world settings. 

In general, various machines can exhibit diverse frequency patterns in their emitted sound, and normal or abnormal sounds can also possess distinct frequency patterns. In practical applications, it is strongly desired for ASD to have the capability to automatically identify the frequency pattern of a machine sound and adjust its processing accordingly based on the specific frequency pattern to attain the most optimal results.

In this paper, we propose an ASD method using self-attention-based frequency pattern analysis (ASD-AFPA) to extract essential information over frequencies of the machine sound for improved anomaly detection. It uses STgram-MFN \cite{liu2022anomalous} as the backbone. However, ASD-AFPA differs from STgram-MFN in that it integrates self-attention mechanism \cite{NIPS2017_3f5ee243} after the Log-Mel converter to achieve a more effective spectral feature, before performing spectral-temporal information fusion. To the best of our knowledge, the proposed method is the first to introduce automatic frequency pattern analysis for anomalous sound detection. Our experiments demonstrate that the self-attention module automatically and adaptively analyses the effective frequencies of a machine sound for ASD and enhances that information in the spectral feature representation, which eventually improves the audio feature obtained from the spectral-temporal information fusion.

\begin{figure*}[htp]
    \centering
    \includegraphics[width=.98\textwidth]{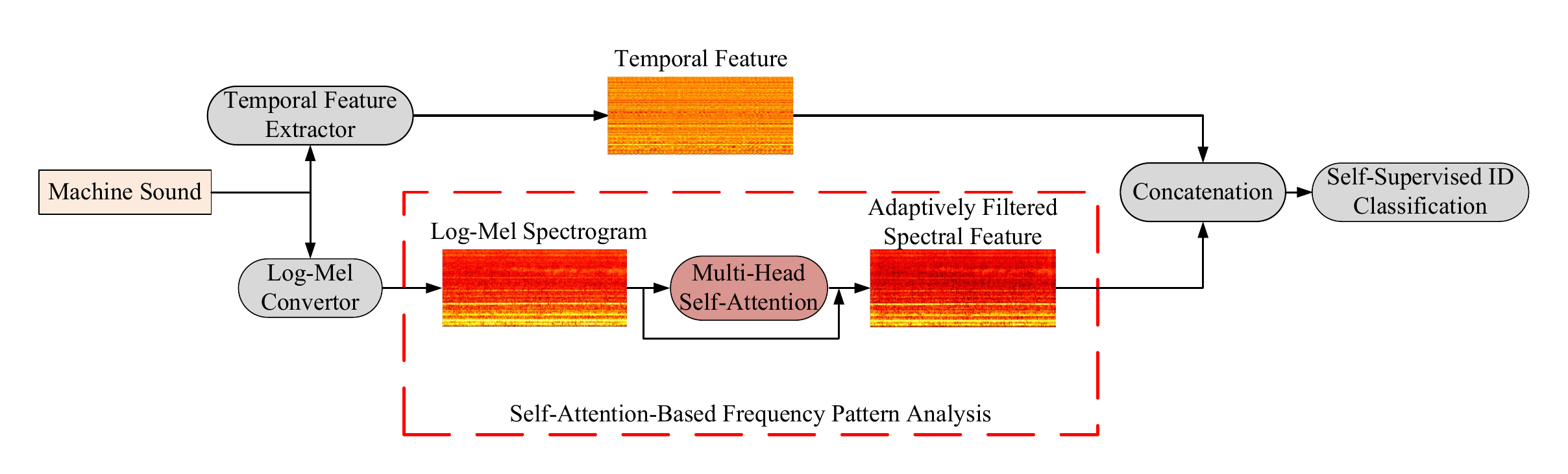}
    \caption{
    Framework of the proposed method with self-attention-based frequency pattern analysis.
    }
    \label{fig:workflow}
\end{figure*}

\section{Proposed Method}

The proposed ASD method using self-attention-based frequency pattern analysis (ASD-AFPA) is illustrated in Figure~\ref{fig:workflow}. The method is built on our previous spectral-temporal information fusion-based ASD approach, STgram-MFN \cite{liu2022anomalous}, but with spectral features boosted using a multi-head self-attention mechanism \cite{NIPS2017_3f5ee243}. This enables the method to automatically detect the frequency pattern of the machine sound and modify its processing accordingly, thereby achieving the most optimal performance for ASD, as demonstrated by the experiment in Section 3. Here, Section 2.1 introduces the STgram-MFN method as the backbone of the proposed method, and Section 2.2 details the multi-head self-attention mechanism enabling the frequency pattern analysis of the machine sound. 

\subsection{Spectral-Temporal Feature Fusion}

For a machine sound as single-channel audio signal $\mathbf{x} \in \mathbb{R}^{1 \times L}$ with length $L$, its Log-Mel spectrogram is $\mathbf{X}_{F} \in \mathbb{R} ^{M \times N} $, where $M$ denotes the Mel bins of the Log-Mel spectrogram (i.e., the number of frequency components) and $N$ denotes the number of time frames. The temporal feature is obtained as 
\begin{equation}
\label{eq:TgramNet}
 \mathbf{X}_{T} =  TN(\mathbf{x}),
\end{equation}
where $TN(\cdot)$ represents TgramNet in \cite{liu2022anomalous}, 
and $ \mathbf{X}_{T} \in \mathbb{R}^{M \times N}$ has the same dimension with $\mathbf{X}_{F} $. The audio representation $\mathbf{X} \in \mathbb{R}^{2\times M \times N}$ through spectral-temporal feature fusion is
\begin{equation}
 {\mathbf{X}} =  Concat_{3D}( \mathbf{X}_{F}, \mathbf{X}_{T}),
\end{equation}
 where $Concat_{3D}(\cdot)$ is a 3-dimensional concatenation operation.

After the feature concatenation, the audio representation $\mathbf{X}$ will be passed to a self-supervised ID classification for anomaly detection, including MobileFaceNet\cite{chen2018mobilefacenets} (MFN) as the classifier and ArcFace\cite{deng2019arcface} as the loss function, which is conducive to enhancing inter-class compactness and amplifying intra-class differences.

Our prior research \cite{liu2022anomalous} has shown that combining spectral and temporal features enhances audio representation for ASD. The proposed method in this paper retains this structure to preserve this benefit.

\subsection{Self-Attention-Based Frequency Pattern Analysis}
The proposed method ASD-AFPA applies the multi-head self-attention (MHSA) mechanism \cite{NIPS2017_3f5ee243} to the Log-Mel spectrogram $\mathbf{X}_{F}$ to automatically analyse the frequency pattern of the machine sound and adaptively using the obtained information of effective frequency components for optimised ASD. 

First,  to prevent information interference between time frames and allows better analysis of the effective frequency components from the Log-Mel spectrogram, we segment the Log-Mel spectrogram over the time dimension as 
\begin{equation}
\label{eq:4}
{\mathbf{X}_{F}} = \left[ {\mathbf{X}_{F}}{(1)},\cdots, {\mathbf{X}_{F}}{(i)}, \cdots, {\mathbf{X}_{F}}{(I)}\right], 
\end{equation}
where ${\mathbf{X}_{F}}{(i)} \in \mathbb{R} ^{M \times {n}}, i = 1, 2, \cdots, I$, and $n = N/I$. Here, $I$ is the number of heads of the multi-head self-attention used for frequency pattern analysis. 

The latent parameters $\mathbf{Q},\mathbf{K},\mathbf{V} \in \mathbb{R} ^{M \times N}$ of the multi-head self-attention mechanism are obtained by the linear mapping of the input Log-Mel spectrogram $\mathbf{X}_{F}$, and they are used to calculate the weights of different frequency components. $\mathbf{Q},\mathbf{K},\mathbf{V}$ are calculated as
\begin{equation}
\label{eq:QKV}
\left\{\begin{matrix} 
  \mathbf{Q} = {\mathbf{X}_{F}} \cdot \mathbf{W}_{Q}, \\  
  \mathbf{K} = {\mathbf{X}_{F}} \cdot \mathbf{W}_{K}, \\
  \mathbf{V} = {\mathbf{X}_{F}} \cdot \mathbf{W}_{V}, 
\end{matrix}\right.  
\end{equation}
where $ \mathbf{W}_{Q},\mathbf{W}_{K},\mathbf{W}_{V} \in \mathbb{R} ^{N \times N} $ are learnable parameter matrices, and their function is to linearly map the input ${\mathbf{X}_{F}}$.

\begin{table*}[!htbp]
\label{tab:1}
\scriptsize
    \centering
	\setlength{\belowcaptionskip}{1pt}
	\setlength{\abovecaptionskip}{1pt}
	\caption{
 Performance on AUC (\%) and pAUC (\%) for different machine types. STgram-MFN is the backbone of the proposed ASD-AFPA method. The proposed method only differs from the backbone in adding the multi-head self-attention to adaptively learn the important frequency patterns for more effective audio feature learning for ASD.
 }
		\resizebox{0.98\textwidth}{!}{
            \begin{tabular}{ccccccccccccccccccccccc}
			\toprule
			\multicolumn{2}{c}{\multirow{2}{*}{Methods}} &\multicolumn{2}{c}{Fan}&\multicolumn{2}{c}{Pump}&\multicolumn{2}{c}{Slider}&\multicolumn{2}{c}{Valve}&\multicolumn{2}{c}{ToyCar}&\multicolumn{2}{c}{ToyConveyor}&\multicolumn{2}{c}{Average} \\
			\cmidrule(r){3-4} \cmidrule(r){5-6} \cmidrule(r){7-8} \cmidrule(r){9-10} \cmidrule(r){11-12} \cmidrule(r){13-14} \cmidrule(r){15-16} 
			\multicolumn{2}{c}{} & {AUC} & {pAUC} & {AUC} & {pAUC} & {AUC} & {pAUC} & {AUC} & {pAUC}& {AUC} & {pAUC} & {AUC} & {pAUC} & {AUC} & {pAUC} \\
			\midrule
			\multicolumn{2}{c}{IDNN\cite{suefusa2020anomalous}} 
& 67.71  & 52.90  & 73.76  & 61.07  & 86.45  & 67.58  & 84.09  & 64.94  & 78.69  & 69.22  & 71.07  & 59.70  & 76.96  & 62.57 \\
			\multicolumn{2}{c}{MobileNetV2\cite{Giri2020a}} 
			& 80.19  & 74.40  & 82.53  & 76.50  & 95.27  & 85.22  & 88.65  & 87.98  & 87.66  & 85.92  & 69.71  & 56.43  & 84.34  & 77.74 \\
		\multicolumn{2}{c}{Glow\_Aff\cite{dohi2021flow}}
		& 74.90  & 65.30  & 83.40  & 73.80  & 94.60  & 82.80  & 91.40  & 75.00  & 92.20  & 84.10 
		& 71.50  & 59.00 
		& 85.20  &73.90 \\
        \midrule
		\multicolumn{2}{c}{{STgram-MFN\cite{liu2022anomalous} (backbone)}} & 94.04  & 88.97  & 91.94  & 81.75  & 99.55  & 97.61  & \textbf{99.64}  & \textbf{98.44}  & 94.44  & 87.68  & 74.57  & 63.60  & 92.36  & 86.34 \\
		\multicolumn{2}{c}{\textbf{ASD-AFPA}} & \textbf{97.55}  & \textbf{93.48}  & \textbf{94.46}  & \textbf{86.76}  & \textbf{99.69}  & \textbf{98.40}  & 99.12  & 95.42  & \textbf{96.15}  & \textbf{89.45}  & \textbf{76.49}  & \textbf{64.21}  & \textbf{93.91}  & \textbf{87.95} \\
			\bottomrule
			\bottomrule
			\end{tabular}
        }
	\label{tab:1}
\end{table*}

To achieve the proposed frequency pattern analysis and enhance the effective frequency components information in the audio feature, ASD-AFPA uses a self-attention mechanism for each part of the segmented input Log-Mel spectrogram to obtain the frequency patterns, which can be calculated as 
\begin{equation}
\label{eq:self_attention}
A({\mathbf{X}_{F}}{(i)}) = softmax \left(\frac{\mathbf{Q}_{i}\cdot \mathbf{K}_{i}^\top  }{\sqrt{n} } \right)\cdot \mathbf{V}_{i},
\end{equation}
\begin{equation}
\label{eq:Attention_map}
\mathbf{D}{i} = softmax \left(\frac{\mathbf{Q}_{i}\cdot \mathbf{K}_{i}^\top  }{\sqrt{n} } \right),
\end{equation}
where $\top$ represents the transposition of the matrix, and $n$ represents the dimension of the time frame of ${\mathbf{X}_{F}{(i)}}$. $\mathbf{Q}_{i}$,  $\mathbf{K}_{i}$, and $\mathbf{V}_{i}$ are part of the latent parameters $\mathbf{Q}$,  $\mathbf{K}$, and $\mathbf{V}$, respectively, corresponding to the spectrogram segment ${\mathbf{X}_{F}}{(i)}$.
Here, $A({\mathbf{X}_{F}}{(i)})$  denotes the output of the self-attention applied on  ${\mathbf{X}_{F}(i)}$, and  $\mathbf{D}{i} \in \mathbb{R} ^{M \times M}$ is the frequency pattern weight matrix (i.e., attention map) of  $\mathbf{X}_{F}{(i)}$ learned from the self-attention mechanism.  Note that the values in $\mathbf{D}{i}$ range from 0 to 1 and represent the importance of the frequency components in the Log-Mel spectrogram. A larger value indicates the weighted frequency components containing more effective information.   

The output of MHSA is $MHSA({{\mathbf{X}_{F}}})$, which is obtained by passing ${\mathbf{X}_{F}}{(i)} (i=1,..., I)$ through the self-attention mechanism and connecting $A({\mathbf{X}_{F}}{(i)})$ on the time frame dimension, that
\begin{equation}
\label{eq:MHSA}
MHSA({\mathbf{X}_{F}}) = concat(A({\mathbf{X}_{F}}{(1)} ) ,\cdots,  A({\mathbf{X}_{F}}{(I)})).  
\end{equation}
To obtain the important information in frequency components while preserving the global information of the Log-Mel spectrogram, we add the residual to the output of MHSA $MHSA({{\mathbf{X}_{F}}})$.  The audio feature $\hat{\mathbf{X}}_{F} \in \mathbb{R} ^{M \times N} $ with adaptively frequency pattern analysis can be calculated as 
\begin{equation}
\label{eq:Residual}
\hat{\mathbf{X}}_{F} = MHSA({\mathbf{X}_{F}}) + {\mathbf{X}_{F}}.
\end{equation}

Finally, the enhanced audio feature representation ${\hat{\mathbf{X}}} \in \mathbb{R}^{2\times M \times N} $ can be obtained by fusing the adaptively filtered spectral feature $\hat{\mathbf{X}}_{F}$ and the temporal feature $\mathbf{X}_{T}$, that
\begin{equation}
{\hat{\mathbf{X}}} =  Concat_{3D}( \hat{\mathbf{X}}_{F}, \mathbf{X}_{T}).
\end{equation}
We adopt the temporal feature to compensate for the possibly missed information in the Log-Mel spectrogram, further improving the audio feature representation with the adaptive frequency pattern analysis.

\section{Experimental Results}
To assess the effectiveness of the proposed method, we performed experiments on the DCASE 2020 Challenge Task 2 dataset \cite{Koizumi2020}. The experimental results demonstrated that incorporating self-attention-based frequency pattern analysis into the existing backbone improved ASD performance compared to state-of-the-art techniques. Furthermore, the ablation study validated the improvement from the proposed self-attention-based frequency pattern analysis. We also present illustrative examples to showcase the important frequency components detected from machine sounds and the resulting changes in the learned spectral features, ultimately leading to improved performance in detecting anomalous sounds.

\subsection{Experimental Setup}
\textbf{Dataset:} We evaluated our proposed method on the DCASE 2020 Challenge Task 2 dataset \cite{Koizumi2020}. The dataset consists of six machine types (Fan, Pump, Slider, Valve, ToyCar, and ToyConveyor), each comprising sounds from four different machine IDs, except for ToyConveyor, which has three different machine IDs. The development and additional datasets' training data is used for training, and the development dataset's test data is used for evaluation. We didn't choose datasets from DCASE 2021 \cite{Kawaguchi_arXiv2021_01} or 2022 \cite{baseline22} since they focus on the domain shift problem, which is out of the scope of this paper. 

\textbf{Evaluation metrics:} The evaluation metrics include the area under the receiver operating characteristic curve (AUC) and the partial-AUC (pAUC), following \cite{suefusa2020anomalous, Giri2020a, dohi2021flow, liu2022anomalous}, where pAUC represents the AUC over a low false-positive-rate range [$0, 0.1$] \cite{Koizumi2020}. A larger metric value indicates a better distinguishing ability for anomalous sound detection.

\textbf{Parameter settings:} We employ Adam optimizer \cite{2014Adam}  with a learning rate of  $1 \times 10{^{-4}} $ for model training by 200 epochs, and cosine annealing is applied for learning rate decay. The margin and scale of the ArcFace loss \cite{deng2019arcface} are empirically set as 1.0 and 30, respectively. The number of heads in Eq.~\eqref{eq:4} is 6. The number of the Mel bins (i.e., frequency components) of the input Log-Mel spectrogram is 128, with 312 time frames.

\subsection{Performance Comparison and Ablation Study}

The proposed method ASD-AFPA is compared with the state-of-the-art methods on the DCASE 2020 dataset, IDNN\cite{suefusa2020anomalous}, MobileNetV2\cite{Giri2020a}, Glow\_Aff\cite{dohi2021flow}, and STgram-MFN\cite{liu2022anomalous}. Table~\ref{tab:1} shows that the proposed ASD-AFPA method significantly improves the ASD performance for all machine types (except Valve), with 1.55\% improvement on AUC and 1.61\% improvement on pAUC, averaged over all the six machine types, compared with STgram-MFN\cite{liu2022anomalous} that achieved the best performance amongst other methods. Note that the STgram-MFN is the backbone of the proposed method, and the only difference between these two methods is that the backbone does not have the self-attention-based frequency pattern analysis, but the proposed ASD-AFPA method does. So this result demonstrated that the proposed method's adaptive frequency pattern analysis (AFPA) is effective for ASD.

\begin{figure*}[!htbp]
    \centering
    \includegraphics[width=\textwidth]{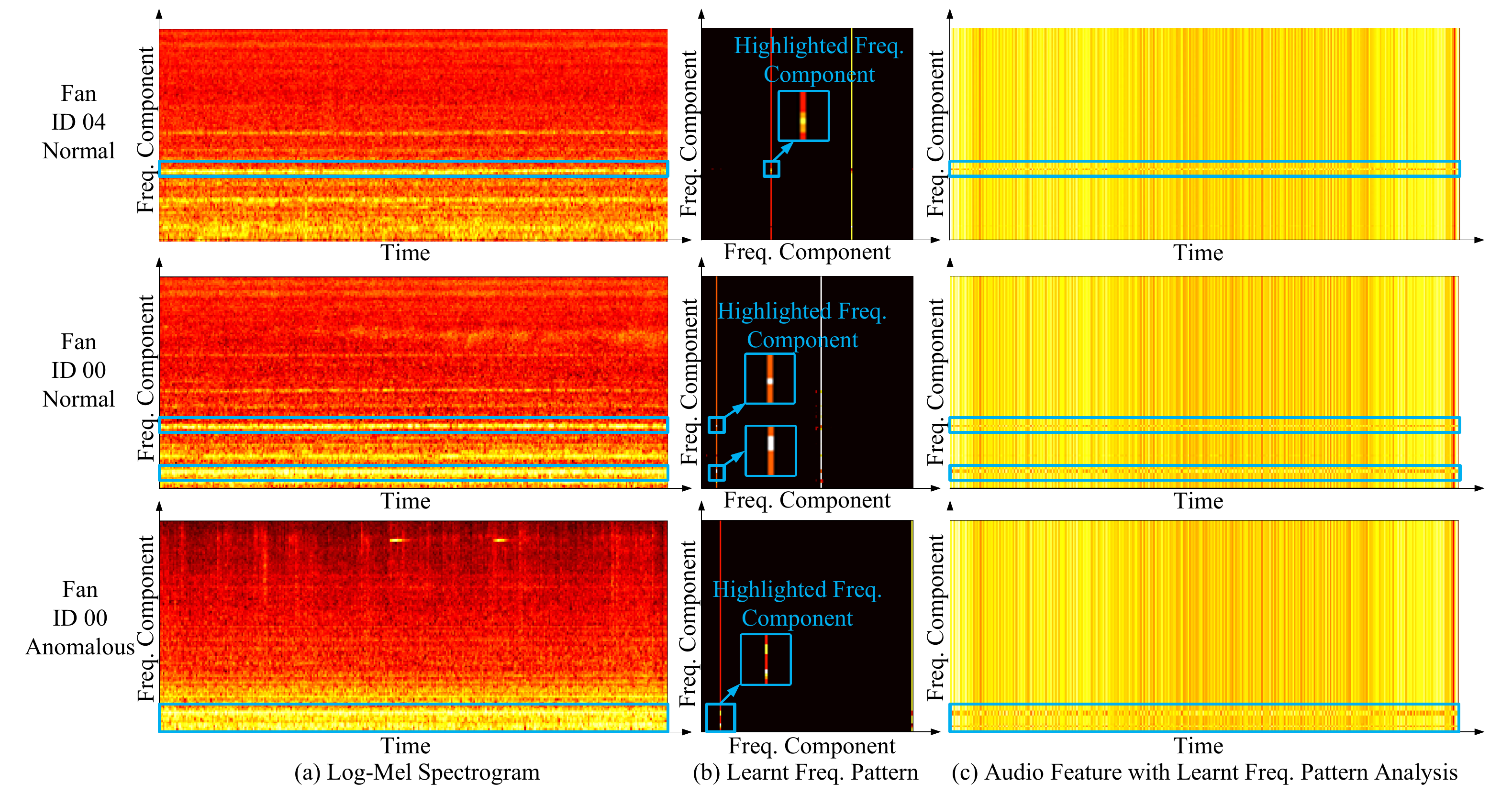}
    \caption{Illustration of the audio feature with the learnt frequency pattern for machine type Fan with ID 04 and 00, respectively. (a) The Log-Mel spectrograms of the input audio signals; (b) The learnt frequency patterns, i.e., the learnt weight matrices corresponding to the input Log-Mel spectrograms of (a);  (c) The learnt audio feature with the proposed self-attention-based frequency pattern analysis. For better understanding, we mark the highlighted important frequency components obtained by the learnt frequency patterns, as shown in the blue boxes of columns (b) and (c).}
 \label{fig:Visualization}
\end{figure*}

\subsection{Visualisation Analysis}

Figure~\ref{fig:Visualization} presents illustrative examples to show the important frequency components detected from machine sounds and the resulting changes in the learned spectral features, ultimately leading to improved performance in detecting anomalous sounds. Figure~\ref{fig:Visualization} includes examples of machine sound from different machine individuals of the same machine type (i.e., ID 00 and ID 04 for Fan), as well as the normal and anomaly sound of the same machine. Column (a) presents the Log-Mel spectrograms of the sound signals. Column (b) presents the learnt frequency patterns, i.e., the mean pooling results of all the learnt frequency weight matrices from Eq. \eqref{eq:Attention_map}. Column (c) presents the learnt audio feature with self-attention-based frequency pattern analysis. 

\begin{figure}[htbp]
    \vspace{-3mm}
    \centering
        \subfloat[STgram-MFN (backbone) \cite{liu2022anomalous}.]{
            \includegraphics[width=\columnwidth]{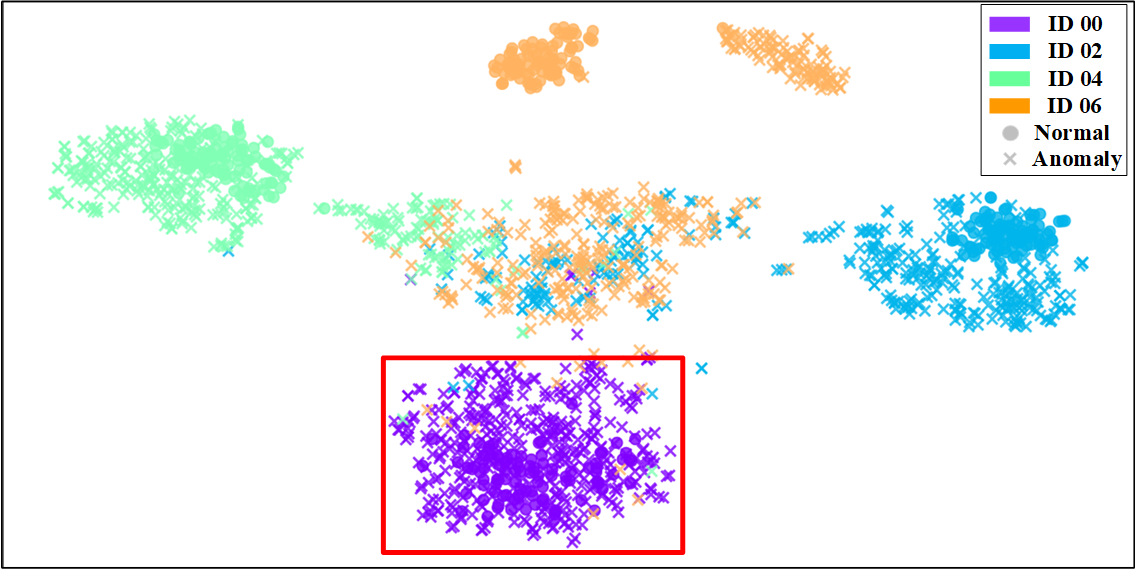}
            \label{subfig:stgram}
        }
        \vspace{-3mm}
        \quad%
        \subfloat[The proposed ASD-AFPA method.]{
            \includegraphics[width=\columnwidth]{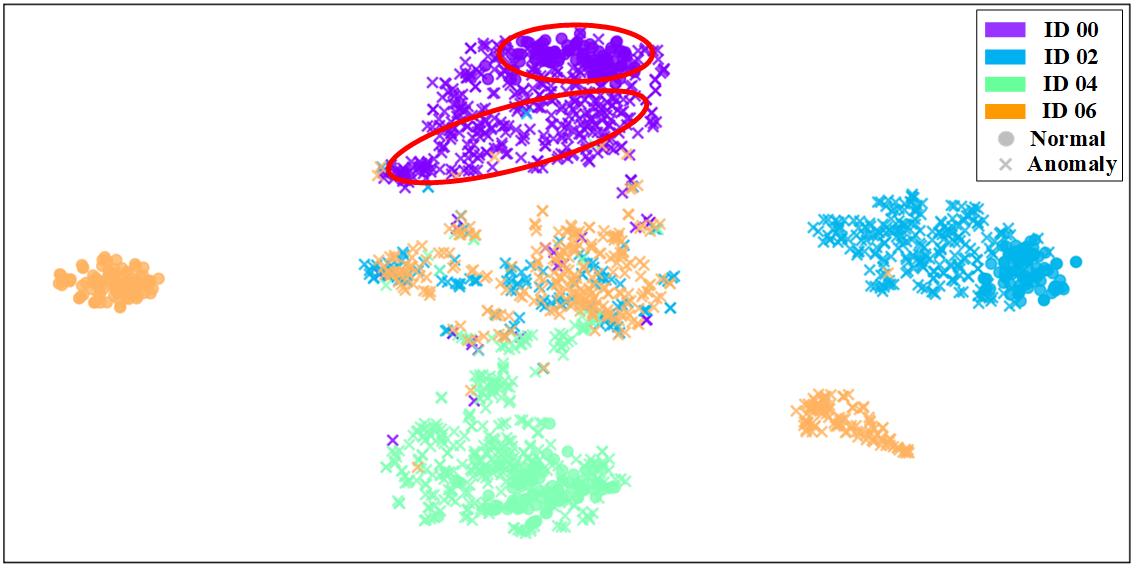}
            \label{subfig:proposed}
        }
        \vspace{-3mm}
    \caption{The t-SNE visualisation comparison between the backbone method STgram-MFN \cite{liu2022anomalous} and the proposed ASD-AFPA for the machine type Fan.}
    \label{fig:tsne}
\end{figure}

From Figure~\ref{fig:Visualization}, we can see that our method can learn different frequency patterns for different operating sound signals,  as demonstrated in the blue boxes in column (b), which can highlight the important frequency components of the input Log-Mel spectrograms, by giving large weights to these frequency components, where the important frequency components with effective information can be highlighted,  as illustrated in column (c). In addition, we can see that our proposed method is effective for all test sound signals, regardless of machine conditions, including signals from different machines (i.e., ID 00 and ID 04 of machine type Fan) and the machine operating status (i.e., normal or abnormal). The results further verify the effectiveness of the proposed method, and show how our method can achieve the adaptive frequency pattern analysis, thus obtaining enhanced audio feature representation to improve the detection performance.

In addition, we provide the t-SNE visualisation comparison between the backbone STgram-MFN \cite{liu2022anomalous} and our proposed ASD-AFPA method for the machine type Fan. As illustrated in Figure~\ref{fig:tsne}, where we can see that our method with the adaptive frequency pattern analysis can further improve the distinguishing ability for anomalous sound detection. The normal and anomalous sound of machine ID 00 can be better distinguished than the backbone method, which further verifies the effectiveness of the proposed method.

\section{Conclusion}

In this paper, we propose an anomalous sound detection method using self-attention-based frequency pattern analysis. It enables automatic detection of the individual frequency pattern of a machine sound and enhances the spectral-temporal feature fusion-based audio feature representation for anomaly detection. Experiments have shown that our proposed approach outperforms existing methods and can identify important frequency components that contribute to enhanced performance in detecting anomalous sounds. 
\section{Acknowledgements}
This work was partly supported by the Natural Science Foundation of Heilongjiang Province under Grant No. YQ2020F010, and a GHfund with Grant No. 202302026860.

\bibliographystyle{IEEEtran}
\bibliography{mybib}

\end{document}